\documentclass[aps,pra,twocolumn,superscriptaddress,showpacs]{revtex4-1}
\usepackage{CJK}
\usepackage[dvips]{graphicx}
\usepackage{graphicx}
\usepackage{amsfonts}
\usepackage{amsmath}
\usepackage{amssymb}
\usepackage{subfigure}

\begin{document}
\begin{CJK*}{GBK}{song}

\title{Quantum discord and its asymptotic behaviors in a time-dependent $XY$ spin chain}

\author{Jian Zhang}
\email{zhangjianphy@gmail.com}
\affiliation{Key Laboratory of Cluster Science of Ministry of Education and School of Chemistry, Beijing Institute of Technology, Beijing 100081, People's Republic of China}
\affiliation{School of Electronic and Information Engineering, Hefei Normal University, Hefei 230601, People's Republic of China}
\author{Bin Shao}
\email{sbin610@bit.edu.cn}
\affiliation{Key Laboratory of Cluster Science of Ministry of Education and School of Chemistry, Beijing Institute of Technology, Beijing 100081, People's Republic of China}

\author{Lian-Ao Wu}
\email{lianao\_wu@ehu.es}
\affiliation{Department of Theoretical Physics and History of Science, The Basque Country University (EHU/UPV),Post Office Box 644, 48080
Bilbao, Spain}
\affiliation{IKERBASQUE, Basque Foundation for Science, 48011, Bilbao, Spain}

\author{Jian Zou}
\affiliation{Key Laboratory of Cluster Science of Ministry of Education and School of Chemistry, Beijing Institute of Technology, Beijing 100081, People's Republic of China}

\date{\today}

\begin{abstract}
We study the dynamics and the asymptotic behaviors of
quantum discord in a one-dimensional $XY$ model coupled
through time-dependent nearest-neighbor interactions and in the presence
of a time-dependent magnetic field. We find that the time evolution of the
nearest-neighbor quantum discord in this system shows
non-ergodic behaviors but is asymptotic to its steady value at the long-time limit.
The zero-temperature asymptotic behaviors of quantum discord
is only determined by the ratio between the coupling parameter and magnetic field, whereas
the finite-temperature asymptotic behaviors determined by both of them.
These asymptotic behaviors are sensitive not only to the initial
values of the coupling parameter and magnetic field, but also to the
final values. It is interesting to note that quantum discords are more robust than entanglement
against the effect of temperature. We also find particular parameter regimes, where the
nearest-neighbor quantum discord is enhanced significantly.

\end{abstract}
\maketitle

\section{Introduction}
Entanglement \cite{RevModPhys.81.865} is a quantum correlation without classical counterpart and has been widely believed as the main reason for the computational advantage of quantum over classical algorithms.
However, it has been found recently that there exist strong indications \cite{PhysRevLett.100.050502} that an more general quantum correlation, quantum discord (QD) \cite{PhysRevLett.88.017901,0305-4470-34-35-315}, is the resource responsible for the speed up in the deterministic quantum computation with one quantum bit (DQC1) \cite{PhysRevLett.81.5672}. Moreover, quantum discord is more robust than the entanglement against
decoherence such that quantum algorithms based only on quantum discord might be more robust than those on entanglement.
Quantum discord \cite{PhysRevLett.88.017901,0305-4470-34-35-315}, as a up-and-coming quantum correlation, often arises as a consequence of coherence between different partitions in a quantum system, being present even in separable states.
This quantum correlation has received much attention in quantum computation \cite{PhysRevLett.100.050502, PhysRevLett.101.200501}, quantum communication \cite{PhysRevA.82.032340,PhysRevA.82.012338, PhysRevA.84.012327}, dynamics of quantum discord \cite{PhysRevA.80.044102,PhysRevA.81.052107, PhysRevLett.105.150501,PhysRevLett.104.200401,PhysRevLett.106.050403}, quantum phase transitions(QPTs) \cite{PhysRevLett.105.095702,PhysRevA.80.022108,PhysRevA.82.042316}, witnessing \cite{PhysRevA.81.062102} etc.

On the other hand, the spin chain model have been proposed as a very reliable model for the future quantum-computing technology in different solid-state systems and a rich model for studying the novel physics of localized spin
systems. This spin chain can be experimentally realized as trapped ions \cite{PhysRevA.63.012306,nphys2252}, superconducting junctions, coupled quantum dots \cite{PhysRevA.57.120,PhysRevB.59.2070}, ultracold quantum gases \cite{nphys2259} et al.
The experimental progresses have triggered intensive theoretical research on  entanglement and quantum discord in the one-dimensional spin chains \cite{PhysRevLett.93.250404,PhysRevLett.98.230503,RevModPhys.80.517,PhysRevA.84.012319,PhysRevA.71.022315}.
Specifically, entanglement and quantum discord in the time-dependent spin chains is investigated \cite{PhysRevLett.107.010403,1742-5468-2011-08-P08026,0295-5075-98-3-30013,HuangPhysRev,Driven_xy_model,Sadiek_2010}.
The dynamics of entanglement in an $XY$ and Ising spin chains has been studied by considering a constant nearest-neighbor coupling and in presence of a time varying magnetic field \cite{HuangPhysRev}. The entanglement dynamics in a time-dependent anisotropic $XY$ model with a small number of spins is studied numerically at zero temperature \cite{Driven_xy_model}. The time-dependent spin-spin coupling was represented by a dc part and a sinusoidal ac part. Recently, Ref. \cite{Sadiek_2010} carefully analyzes the time evolution of entanglement in a one-dimensional spin chain in presence of a time dependent magnetic field $h(t)$ and by considering a time dependent coupling parameter $J(t)$. Both $h(t)$ and $J(t)$ are step functions of time \cite{Sadiek_2010}.  The entanglement undergoes a nonergodic behavior. The zero-temperature asymptotic behaviors of entanglement depend only on the ratio between the coupling parameter and magnetic field, whereas the finite-temperature asymptotic behaviors depend on both the coupling parameter and magnetic field. The aim of this work is to discuss what the advantages of quantum discord is in the time-dependent spin chain, comparing with entanglement and to explore the relation between the asymptotic behaviors of quantum discord and the parameter setting.

This paper attempts to study the dynamics of quantum discord in the time-dependent $XY$ model, and to explore the interesting asymptotic behaviors of quantum discord. In Sec. II, we introduce the time-dependent $XY$ spin chain and describe the general solution for this model. Sec. III analyzes the discord dynamics.  The asymptotic behaviors of quantum discord are analyzed carefully in Sec. III. We conclude our work in Sec. IV.

\section{QUANTUM DISCORD IN THE TIME DEPENDENT XY MODEL}

In this section, we briefly review the exact solution for the spin $XY$ model of a one-dimensional lattices with $N$ sites coupled through time-dependent couplings $J(t)$ and subject to an external time-dependent magnetic field $h(t)$, and introduce the geometric measure of quantum discord.

The Hamiltonian for such a system is given by \cite{Sadiek_2010} (setting $\hbar=1$)
\begin{equation}
H=-\frac{J(t)}{2} \sum_{i=1}^{N}( (1+\gamma)\sigma_{i}^{x} \sigma_{i+1}^{x}+(1-\gamma)\sigma_{i}^{y} \sigma_{i+1}^{y})- h(t)\sum_{i=1}^{N}\sigma_{i}^{z},
\label{eq:H}
\end{equation}
where $\sigma^{\alpha}_{i},\alpha=\{x,y,z\}$ are the Pauli matrices and $\gamma$ is the anisotropy parameter.

The coupling and magnetic field are represented respectively by
\begin{eqnarray}
\nonumber J(t)&=&J_0 + (J_1 - J_0) \theta(t) \\
 h(t)&=&h_0 + (h_1 - h_0) \theta(t),
\end{eqnarray}
where $\theta(t)$ is the mathematical step function
\begin{equation}
\theta(t)=\left\{
\begin{array}{lr}
0 & \qquad t\leq 0 \\
1 & \qquad t>0
\end{array}.
\right.
\end{equation}
We assume that the system is initially in thermal equilibrium. The reduced two-spin density matrix $\rho^{ij}(t)$ for this system therefore is
\begin{equation}
\rho^{ij}(t)=\left(\begin{array}{cccc}
 \rho_{11} &0 &0 & \rho_{14} \\
 0 & \rho_{22} & \rho_{23} &0 \\
 0 & \rho_{23}^{\ast} & \rho_{33} &0 \\
  \rho_{14}^{\ast} &0 &0 &\rho_{44} \\
\end{array}\right)\, ,
\label{eq:rhot}\end{equation}
where matrix elements can be written in terms of one- and two-point correlation functions \cite{Sadiek_2010}
\begin{eqnarray}
\nonumber \rho_{11}&=&\left\langle M^z_{l}\right\rangle+\left\langle S^{z}_{l} S^{z}_{m} \right\rangle+\frac{1}{4} ,\\
\nonumber \rho_{22}&=&\rho_{33}=-\left\langle S^{z}_{l} S^{z}_{m} \right\rangle+\frac{1}{4} ,\\
\nonumber \rho_{44}&=&- \left\langle M^z_{l}\right\rangle+\left\langle S^{z}_{l} S^{z}_{m} \right\rangle+\frac{1}{4} ,\\
\nonumber \rho_{23}&=&\left\langle S^{x}_{l} S^{x}_{m} \right\rangle+\left\langle S^{y}_{l} S^{y}_{m} \right\rangle,\\
\rho_{14}&=&\left\langle S^{x}_{l} S^{x}_{m} \right\rangle-\left\langle S^{y}_{l} S^{y}_{m} \right\rangle .
\end{eqnarray}
The magnetization in the $z$-direction is defined as
\begin{equation}
M^{z}=\frac{1}{N}\sum_{j=1}^{N}(S_{j}^{z}).
\end{equation}
Its expectation value is
\begin{equation}
\left\langle M^{z}\right\rangle=\frac{Tr[M^{z}\rho(t)]}{Tr[\rho(t)]},
\end{equation}
specifically \cite{Sadiek_2010},
\begin{eqnarray}
\nonumber \left\langle M^{z}\right\rangle= \frac{1}{4N}\sum_{p=1}^{N/2}\frac{\tanh[\beta \Gamma(h_{0},J_{0})]}{\Gamma^2(h_{1},J_{1})\Gamma(h_{0},J_{0})}\\
\nonumber \biggl\{2 J_{1}(J_{0} h_{1} -J_{1} h_{0}) \delta_{p}^2 \sin^2[2t \Gamma(h_{1},J_{1})]\\
+4\Gamma^2(h_{1},J_{1})(J_{0}\cos\phi_{p}+h_{0})\biggr\}\, .
\end{eqnarray}
where $\phi_{p}=\frac{2 \pi p}{N}$ , $\delta_{p}=2 \gamma \sin \phi_{p}$ and $\beta=1/k T$. $k$ is Boltzmann constant and $T$ is the temperature.

Using Wick Theorem \cite{Wick}, the nearest-neighbor spin correlation functions can be obtained as
\begin{eqnarray}
\nonumber \left\langle S^{x}_{l} S^{x}_{l+1} \right\rangle&=&\frac{1}{4}F_{l,l+1}\\
\nonumber \left\langle S^{y}_{l} S^{y}_{l+1} \right\rangle&=&\frac{1}{4}F_{l+1,l}\\
\nonumber \left\langle S^{z}_{l} S^{z}_{l+1} \right\rangle&=&\frac{1}{4} \{F_{l, l}\times F_{l+1,l+1}-Q_{l,l+1}\times G_{l,l+1}\\
&-&F_{l+1,l}\times F_{l,l+1}\},
\end{eqnarray}
where \cite{Sadiek_2010}

\begin{widetext}
\begin{eqnarray}
\nonumber Q_{l, m}&=& \frac{1}{N} \sum_{p=1}^{N/2} \biggl\{2\cos[(m-l)\phi_{p}]\\=
&+&\frac{i(J_{1}h_{0}- J_{0}h_{1}) \delta_{p} \sin[(m-l)\phi_{p}]\sin[4t\Gamma(h_{1},J_{1})]\tanh[\beta \Gamma(h_{0},J_{0})]}{\Gamma(h_{1},J_{1})\Gamma(h_{0},J_{0})}\biggr\}\, ,\quad
\end{eqnarray}
\begin{eqnarray}
\nonumber G_{l, m}&=&\frac{1}{N} \sum_{p=1}^{N/2} \biggl\{-2\cos[(m-l)\phi_{p}]\\
&+&\frac{i(J_{1}h_{0}- J_{0}h_{1}) \delta_{p} \sin[(m-l)\phi_{p}]\sin[4t\Gamma(h_{1},J_{1})]\tanh[\beta \Gamma(h_{0},J_{0})]}{\Gamma(h_{1},J_{1})\Gamma(h_{0},J_{0})}\biggr\}\, ,\quad
\end{eqnarray}
\begin{eqnarray}
\nonumber F_{l, m}&=&\frac{1}{N} \sum_{p=1}^{N/2} \frac{\tanh[\beta \Gamma(h_{0},J_{0})]}{\Gamma^2(h_{1},J_{1})\Gamma(h_{0},J_{0})} \Biggl\{\cos[(m-l)\phi_{p}] \\
\nonumber &\times&\biggl\{J_{1} [J_{0} h_{1} - J_{1} h_{0}] \delta^2_{p} \sin^2[2t\Gamma(h_{1},J_{1})] + 2\Gamma^2(h_{1},J_{1})(J_{0}\cos\phi_{p}+h_{0})\biggr\}\\
 &+& \delta_{p} \sin[(m-l)\phi_{p}]\biggl\{J_{0}\Gamma^2(h_{1},J_{1}) +2 (J_{1} h_{0} - J_{0} h_{1}) (J_{1}\cos\phi_{p}+h_{1}) \sin^2[2t\Gamma(h_{1},J_{1})]\biggr\}\Biggr\}\, .
\end{eqnarray}
\end{widetext}
Here
\begin{equation}
\Gamma[h(t),J(t)]=\left\{[J(t)\cos \phi_{p} + h(t)]^{2}+\gamma^2 J^2(t) \sin^2\phi_{p}\right\}^{\frac{1}{2}} .
\end{equation}

Various measures of quantum discord and their extensions to multipartite systems have been proposed.
Here we use the geometric measure of quantum discord \cite{PhysRevLett.105.190502},
\begin{equation}\label{GMD}
D(\rho)=\mathrm{min}_{\chi\in\Omega_0}||\rho-\chi||^2,
\end{equation}
where $\Omega_0$ denotes the set of zero-discord states and
$||X-Y||^2=\mathrm{Tr}(X-Y)^2$ is the square norm in the
Hilbert-Schmidt space. In order to calculate this quantity
for an arbitrary two-qubit state,
we write $\rho$ in bloch representation:
\begin{equation}
\rho=\frac{1}{4}(\openone\otimes\openone+\sum_{i=1}^3x_i\sigma_i\otimes\openone+\sum_{i=1}^3y_i\openone\otimes\sigma_i+
\sum_{i,j=1}^3T_{ij}\sigma_i\otimes\sigma_j),
\end{equation}
where $x_i=\mathrm{Tr}\{\rho(\sigma_i\otimes\openone)\}$,
$y_i=\mathrm{Tr}\{\rho(\openone\otimes\sigma_i)\}$ are components of the
local Bloch vectors,
$T_{ij}=\mathrm{Tr}\{\rho(\sigma_i\otimes\sigma_j)\}$ are components of
the correlation tensor, and $\sigma_i$ ($i \in \{1,2,3\}$) are the
three Pauli matrices.  Each state $\rho$ can be expressed by the parameter set
$\{\vec{x},\vec{y},T\}$, the geometric measure of quantum discord is therefore given explicitly \cite{PhysRevLett.105.190502}
\begin{equation}
D(\rho)=\frac{1}{4}(||\vec{x}||^2+||T||^2-k_{\mathrm{max}}),
\end{equation}
where $k_{\mathrm{max}}$ is the largest eigenvalue of matrix
$K=\vec{x}\vec{x}^{\mathrm{T}}+TT^{\mathrm{T}}$.

\section{Dynamics of quantum discord}
We now come to study the dynamics of quantum discord in the time-dependent $XY$ spin chain.  Throughout the paper we use $N=1000$ to numerically demonstrate our general results, which are almost size-independent. We also employ three dimensionless parameters $\lambda=J/h$, $\lambda_{1}=J_{1}/h_{1}$ and $\lambda_{0}=J_{0}/h_{0}$ for convenience.

The $XY$ model undergoes a second-order QPT (Ising transition) at the critical point (CP) $\lambda_{c} = 1$, which separates a ferromagnetic ordered phase from a quantum paramagnetic phase. When $\lambda> 1$ it is claimed \cite{Bunder99} there is another second-order QPT at $\gamma_{c}= 0$, which is termed as the anisotropy transition. Differently from the Ising transition due to the external field, this transition is driven by the anisotropy parameter $\gamma$ and separates a ferromagnet ordered along the $x$ direction from a ferromagnet ordered along the $y$ direction.

\begin{figure}
 \subfigure{\label{(t)-T0-gamma1}\includegraphics[width=8cm]{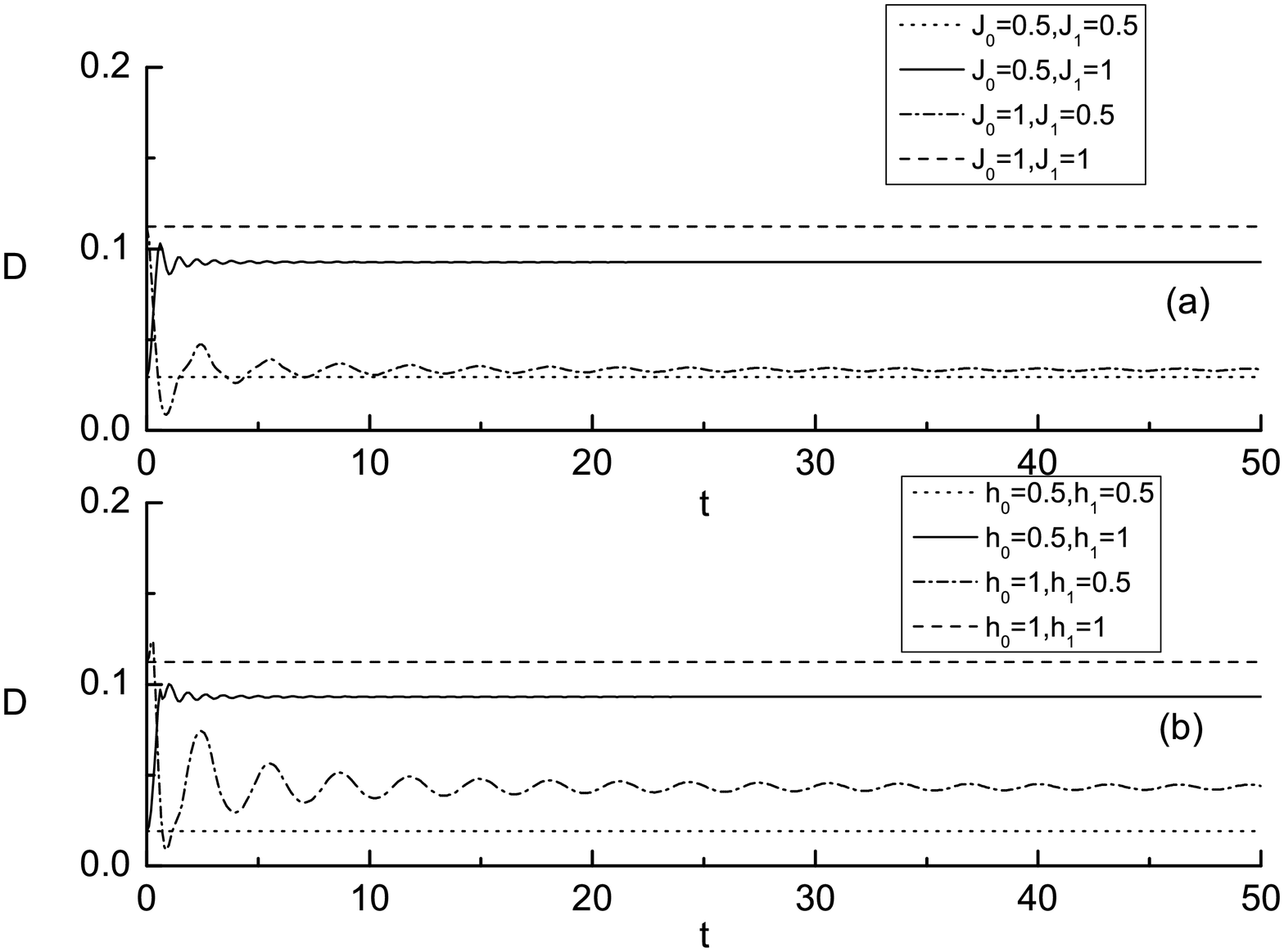}}
	\caption{Dynamics of the nearest-neighbor discord $D(i,i+1)$ with $\gamma=1$, $kT=0$ and (a) $h_{0}=h_{1}=1$ for various values of $J_{0}$ and $J_{1}$; (b) $J_{0}=J_{1}=1$ for various values of $h_{0}$ and $h_{1}$. }\label{(t)-T0-gamma1}
\end{figure}

Fig.~\ref{(t)-T0-gamma1} plots the dynamics of the nearest-neighbor discord $D(i,i+1)$ at zero temperature for the transverse Ising model with $\gamma=1$.
When the coupling parameter (and the magnetic field) is a step function, the discord reaches a value that is neither its value at $J=J_{0}$ ($h=h_{0}$) nor at $J_{1}$ ($h=h_{1}$). In other words, $D(i,i+1)$ shows a non-ergodic behavior, which is similar to that of entanglement \cite{Sadiek_2010}.
At higher temperature quantum discord remains finite as $t \rightarrow \infty$, though the magnitude of the asymptotic discord decreases.
Similarly, $D(i,i+1)$ also shows the non-ergodic behavior in the partially anisotropic $XY$ model with $\gamma=0.5$, though the equilibrium time is longer than that in the Ising model.
On the contrary, $D(i,i+1)$ in the isotropic $XY$ model ($\gamma=0$) becomes a constant determined by the values of $J_{0}$ and $h_{0}$. The isotropy of the initial coupling parameters may make spins equally aligned into the $x$ and $y$ directions, apart from those in the $z$-direction, which results in finite quantum discord. Increasing the coupling parameters strength would not change the associated discord.
\begin{figure}[htbp]
 \centering
   \subfigure{\label{(t,lambda1)-T0-gamma1-J0[1]}\includegraphics[width=7cm]{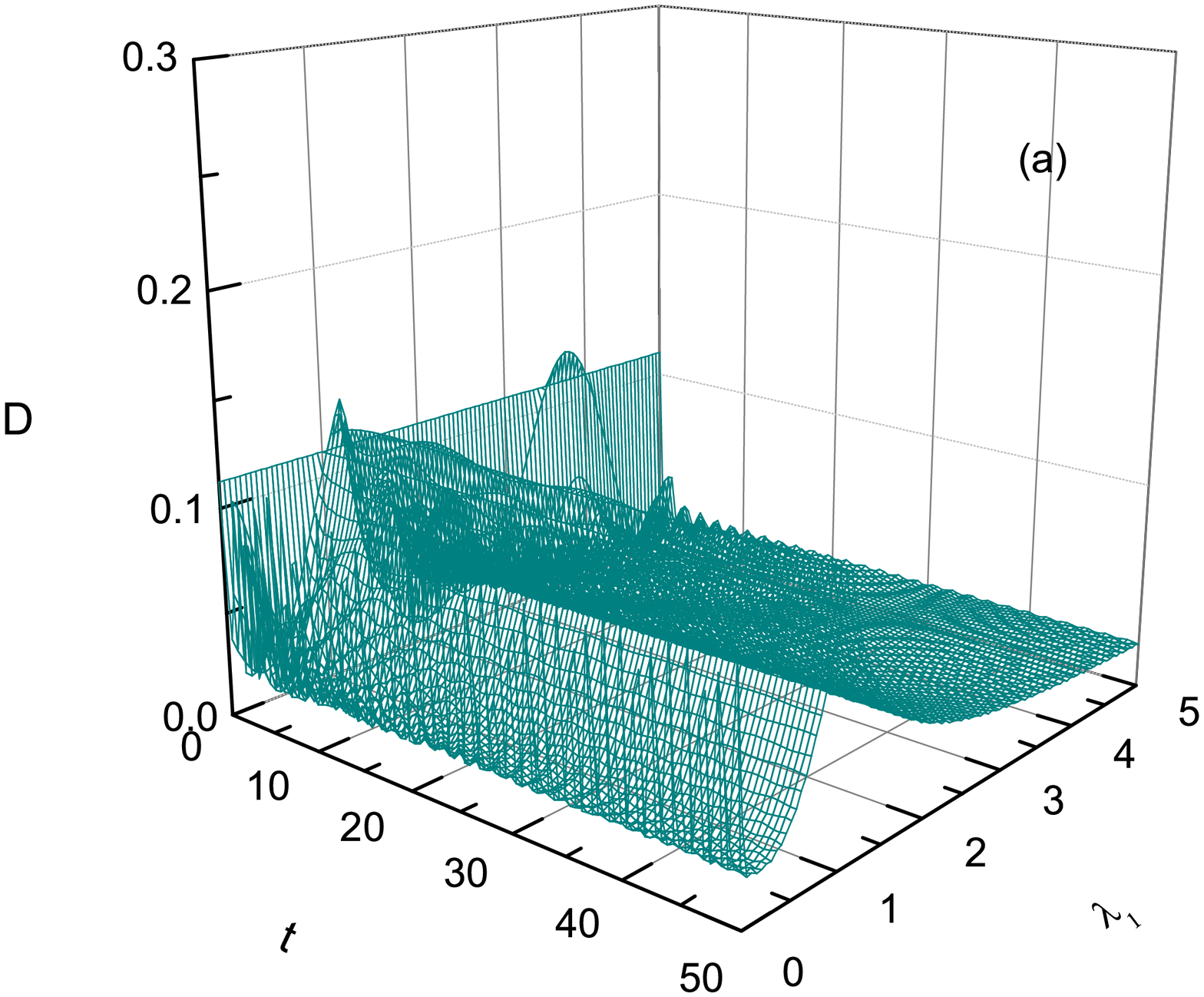}}\quad
   \subfigure{\label{(t,lambda1)-T0-gamma1-J0[5]}\includegraphics[width=7cm]{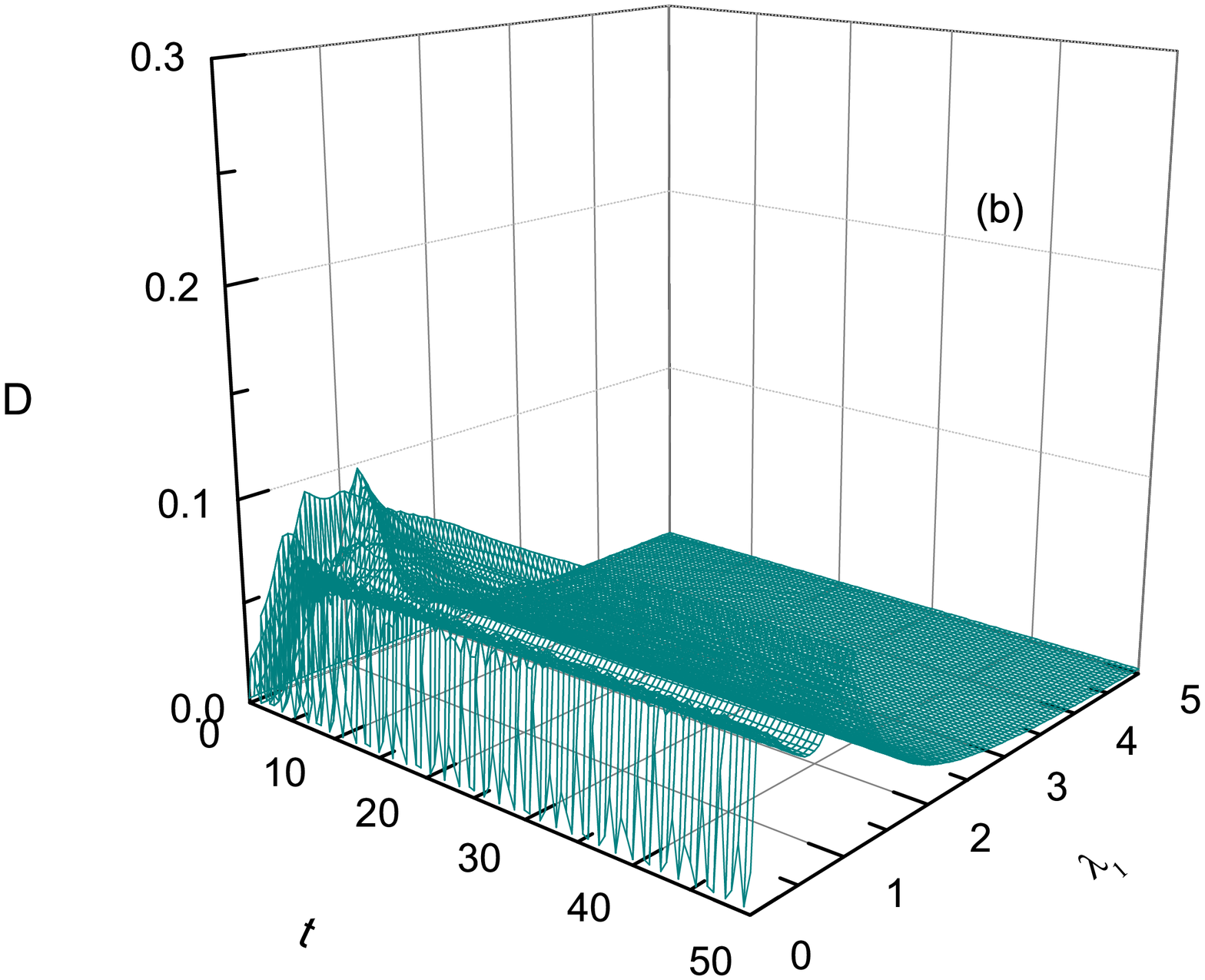}}
   \caption{(Color online)$D(i,i+1)$ as a function of $\lambda_{1}$ and $t$ at $kT=0$ with $\gamma=1$, $h_{0}=h_{1}=1$ and (a) $J_{0}=1$; (b)$J_{0}=5$.}\label{(t,lambda1)-T0-gamma1}
\end{figure}

We now focus on the dynamics of quantum discord as a function of $\lambda_1$.
In Fig.~\ref{(t,lambda1)-T0-gamma1}, we plot $D(i,i+1)$ as a function of $t$ and $\lambda_1$  for the Ising model where $h_0=h_1=1$ and $T=0$.
We first consider the case when the system is initially prepared in a state with $\lambda_0=\lambda_c$, i.e., $J_{0}=1$, in Fig.~\ref{(t,lambda1)-T0-gamma1-J0[1]}. When $\lambda_1=0$, the discord oscillates in time. The magnitude of the discord increases with $\lambda_1$ until it reaches its maximum at $\lambda_c$. Interestingly, when $\lambda_1$ exceeds $\lambda_c$, quantum discord behaves differently from that of entanglement shown in \cite{Sadiek_2010}. The discord decreases and becomes steady when $J$ dominates over $h$, while entanglement does not. On the other hand, when the system initially is in the parameter region $\lambda_0 > \lambda_c$, the maximum of quantum discord is much smaller as shown in Fig.~\ref{(t,lambda1)-T0-gamma1-J0[5]} where  $J_{0}=5$.
We see that there are two peaks, and the second is higher and decreases with increase in $\lambda_0$ .
For the partially anisotropic cases, quantum discord behaves similar to that in the Ising case
but with smaller magnitudes. Quantum discord of the completely isotropic $XY$ system is independent of $\lambda_1$.

\section{Asymptotic Behaviors of Quantum Discord}

\begin{figure}[htbp]
 \centering
 \subfigure{\label{(lambda)-gamma1}\includegraphics[width=10cm]{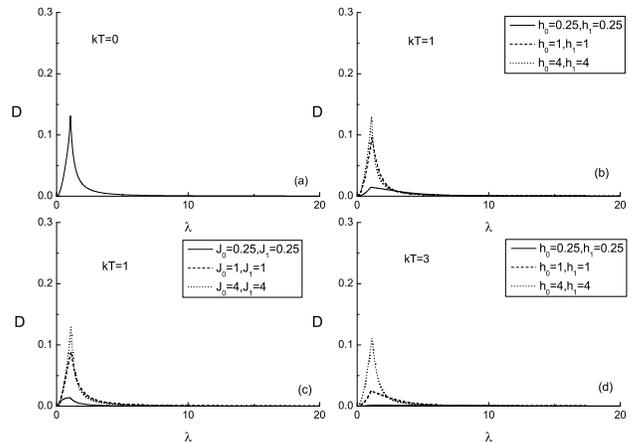}}
\caption{$D(i,i+1)$ as functions of $\lambda$ for $h=h_{0}=h_{1}$ and $J=J_{0}=J_{1}$ at (a) $kT=0$ with different $J$ and $h$; (b) $kT=1$ with $h_{0}=h_{1}=0.25,1,4$; (c)$kT=1$ with $J_{0}=J_{1}=0.25,1,4$; (d) $kT=3$ with $h_{0}=h_{1}=0.25,1,4$ with $\gamma=1$.}
 \label{(lambda)-gamma1}
\end{figure}


We will examine the asymptotic behaviors of the quantum discord $D(i,i+1)$ as a function of $\lambda$ at $(t\rightarrow\infty)$ for different values of $J=J_0=J_1$ and $h=h_0=h_1$ and at different temperatures.  First, for the Ising model ($\gamma=1$), Fig.~\ref{(lambda)-gamma1}(a) shows that the zero-temperature behavior of $D(i,i+1)$ depends only on the ratio $J/h$ ($=\lambda$) rather than their individual values.  $D(i,i+1)$ starts at zero, reaches a maximum and then vanishes at larger values of $\lambda$. The maximum of asymptotic value of $D(i,i+1)$ decreases as the temperature increases as shown in Fig.~\ref{(lambda)-gamma1}(b) and(d).  On the other hand, the finite-temperature discord $D(i,i+1)$ is not only determined by the ratio of $J$ and $h$ but also individual values of $J$ and $h$.  Fig.~\ref{(lambda)-gamma1}(b) and \ref{(lambda)-gamma1}(c) indicate that an increase in $h$ and $J$ causes the maximal discord to increase when $kT=1$.

\begin{figure}[htbp]
 \centering
 \subfigure{\label{(lambda)-gamma0.5}\includegraphics[width=10cm]{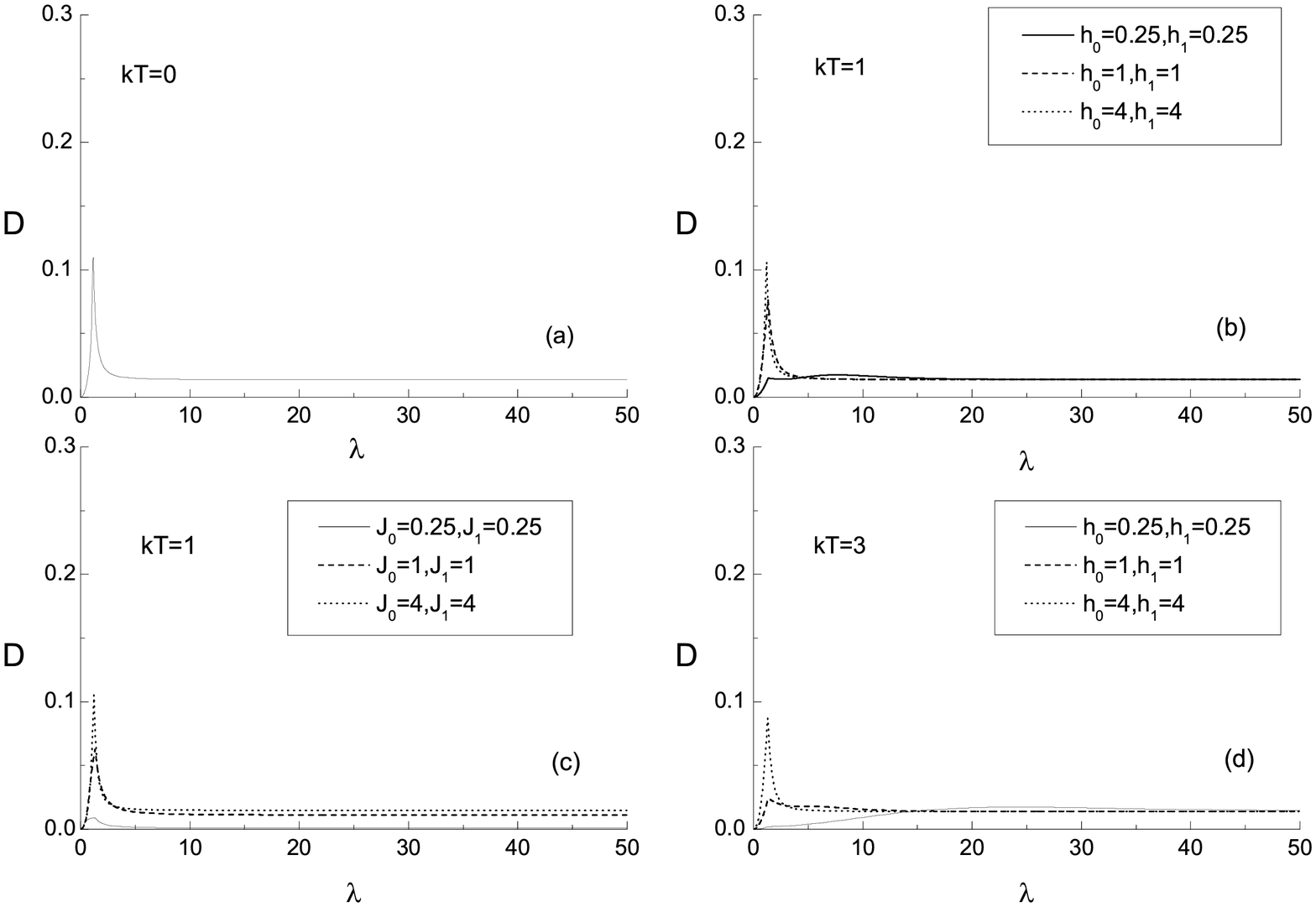}}
\caption{$D(i,i+1)$ as a function of $\lambda$ for $h=h_{0}=h_{1}$ and $J=J_{0}=J_{1}$ at (a) $kT=0$ with different combinations of $J$ and $h$; (b) $kT=1$ with $h_{0}=h_{1}=0.25,1,4$; (c) $kT=1$ with $J_{0}=J_{1}=0.25,1,4$; (d)$kT=3$ with $h_{0}=h_{1}=0.25,1,4$ with $\gamma=0.5$.}
 \label{(lambda)-gamma0.5}
\end{figure}

Next, we discuss the partially anisotropic $XY$ model with $\gamma=0.5$.  Similarly to the Ising case, the zero-temperature discord $D(i,i+1)$ depends only on the ratio $J/h$ as in Fig.~\ref{(lambda)-gamma0.5}(a). $D(i,i+1)$ starts from zero, reaches a maximal value and then decays
to a constant value for larger $\lambda$'s, where entanglement vanishes \cite{Sadiek_2010}. When $h \gg J$, the magnetic field dominates such that spins are aligned along the $z$ direction and the nearest-neighbor discord is zero. When $h \ll J$, the coupling will dominate. For the partially anisotropic model, the strong nearest-neighbor couplings $J$ make the spins aligned isotropically and consequently the discord maintains an equilibrium and finite value ( see also the results for entanglement \cite{Sadiek_2010}).
The critical behavior of quantum discord around the critical point $\lambda=1$ changes considerably as the temperature and other parameters.  The maximal discord is enhanced at high magnetic fields and the stronger coupling $J$.
The temperature effect is shown in Fig.~\ref{(lambda)-gamma0.5}(d). The critical behavior of quantum discord disappears for some values of $h$ and $J$. It manifests that the thermal excitations suppress quantum effect and even ruin the critical behaviors of quantum discord.

\begin{figure}[htbp]
 \centering
 \subfigure{\label{(lambda)-gamma0}\includegraphics[width=10cm]{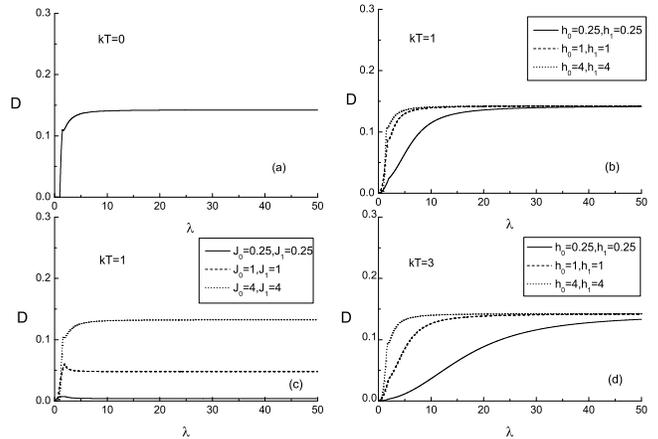}}
\caption{$D(i,i+1)$ as a function of $\lambda$ for $h=h_{0}=h_{1}$ and $J=J_{0}=J_{1}$ at (a)$kT=0$ with different combination of $J$ and $h$; (b)$kT=1$ with $h_{0}=h_{1}=0.25,1,4$;(c)$kT=1$ with $J_{0}=J_{1}=0.25,1,4$; (d)$kT=3$ with $h_{0}=h_{1}=0.25,1,4$ with $\gamma=0$.}
 \label{(lambda)-gamma0}
\end{figure}


Likewise, $D(i,i+1)$ in the isotropic $XY$ model depends only on the ratio $J/h$ at $T=0$, as shown in Fig.~\ref{(lambda)-gamma0}(a). Differently, $D(i,i+1)$ starts from zero and saturates at $\lambda=8$. There seems to be a tiny peak at $\lambda=1$ before the saturation. Interestingly, raising the temperature delays the saturations of the discord but does not reduce their amplitudes, as shown in Fig.~\ref{(lambda)-gamma0}(b). The magnetic fields individually affect the saturations. The smaller the magnetic fields are, the later the saturations appear. In Fig.~\ref{(lambda)-gamma0}(c), the peaks at critical point become visible when increasing $J$ and noticeably
the saturations of $D(i,i+1)$ happen with larger amplitudes. The higher temperatures delay the saturations of the quantum discord as in Fig.~\ref{(lambda)-gamma0}(d).

\begin{figure*}[htbp]
\centering
   \subfigure{\label{[J(),h()]-T0-gamma1}\includegraphics[width=15cm]{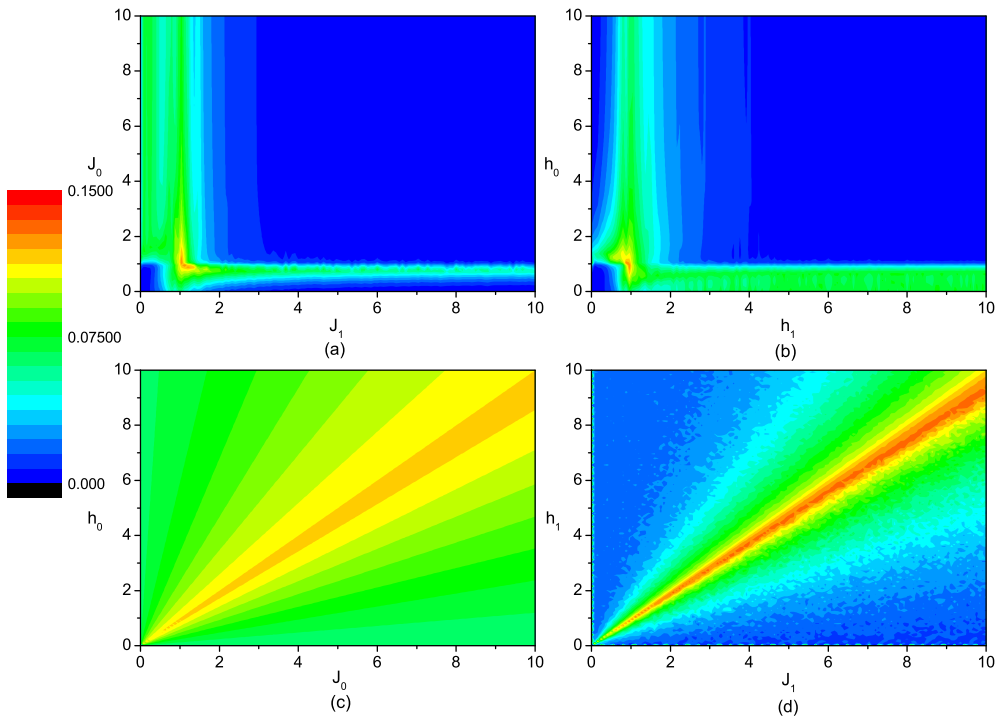}}
   \caption{(Color online)The asymptotic behavior of $D(i,i+1)$ as a function of (a) $J_{0}$ and $J_{1}$ with $h_{0}=h_{1}=1$,(b)$h_{0}$ and $h_{1}$  with $J_{0}=J_{1}=1$, (c) $h_{0}$ and $J_{0}$  with  $h_{1}=J_{1}=1$ and (d) $h_{1}$ and $J_{1}$ with  $h_{0}=J_{0}=1$   at $kT=0$ with $\gamma=1$       .}
 \label{[J(),h()]-T0-gamma1}
\end{figure*}

\begin{figure*}[htbp]
\centering
   \subfigure{\label{[J(),h()]-T0-gamma0.5}\includegraphics[width=15cm]{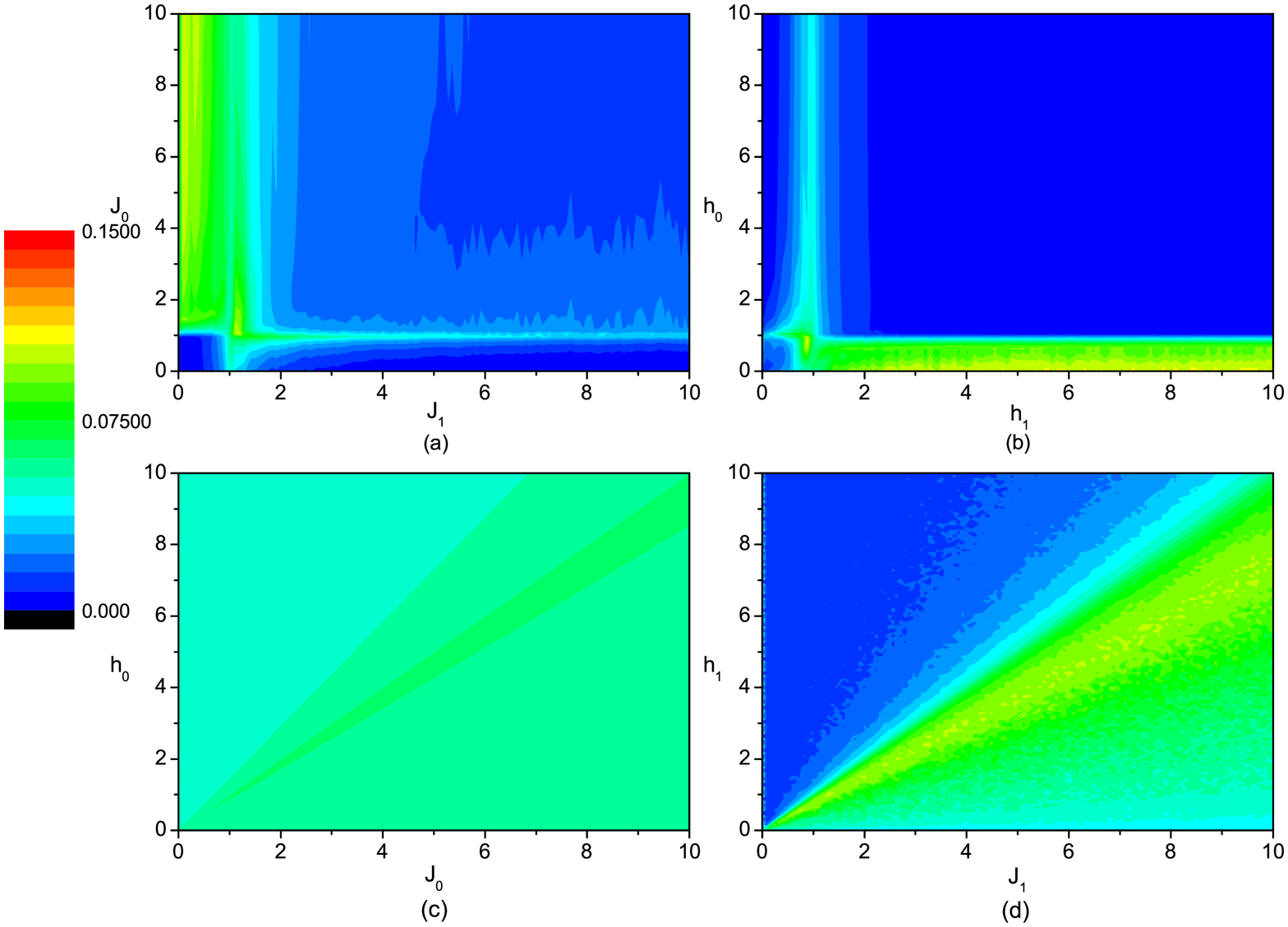}}
   \caption{(Color online)The asymptotic behavior of $D(i,i+1)$ as a function of (a) $J_{0}$ and $J_{1}$ with  $h_{0}=h_{1}=1$, (b)$h_{0}$ and $h_{1}$ with $J_{0}=J_{1}=1$, (c) $h_{0}$ and $J_{0}$ with $h_{1}=J_{1}=1$ and (d) $h_{1}$ and $J_{1}$ with $h_{0}=J_{0}=1$ at $kT=0$ with $\gamma=0.5$.}
 \label{[J(),h()]-T0-gamma0.5}
\end{figure*}
%

\begin{figure*}[htbp]
\centering
   \subfigure{\label{[J(),h()]-T0-gamma0}\includegraphics[width=15cm]{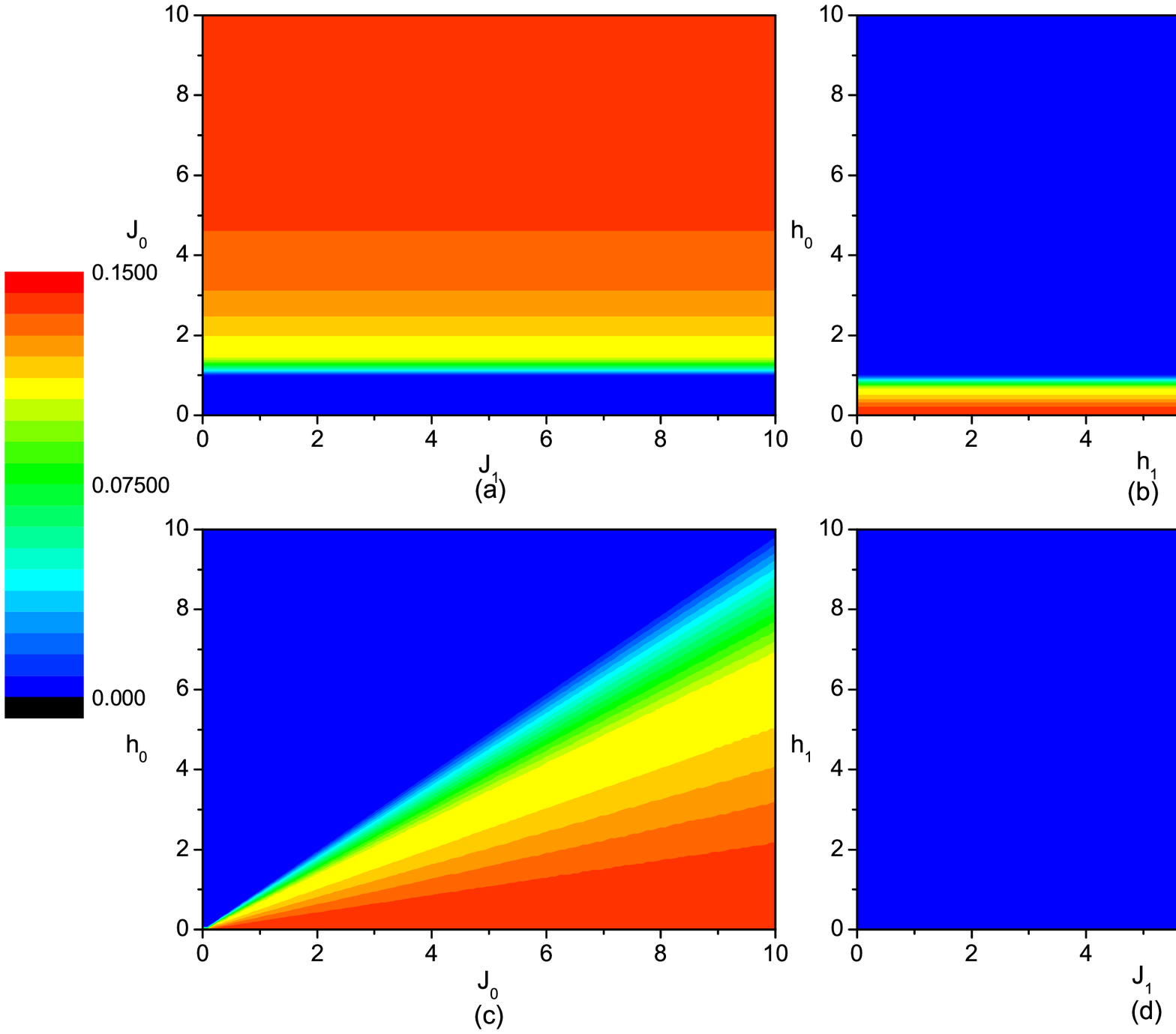}}
   \caption{(Color online)The asymptotic behaviours of $D(i,i+1)$ as a function of (a) $J_{0}$ and $J_{1}$ with $h_{0}=h_{1}=1$, (b)$h_{0}$ and $h_{1}$  with  $J_{0}=J_{1}=1$, (c) $h_{0}$ and $J_{0}$ with  $h_{1}=J_{1}=1$ and (d) $h_{1}$ and $J_{1}$ with $h_{0}=J_{0}=1$ at $kT=0$.}
 \label{[J(),h()]-T0-gamma0}
\end{figure*}

To further study the roles of the magnetic field and coupling parameters $h_0$, $h_1$, $J_0$ and $J_1$, we first calculate the zero-temparature asymptotic behaviors of quantum discord at $t \rightarrow \infty$ in wider parameter regimes.
Fig.~\ref{[J(),h()]-T0-gamma1} has four discord contours for the Ising model at $t \rightarrow \infty$. We plot the discord contour function of $J_{0}$ and $J_{1}$ in Fig.~\ref{[J(),h()]-T0-gamma1}(a), where $h_0=h_1=1$. The discord starts with a zero value for $J_0=J_1=0$, reaches the maximum at $J_{0} \approx J_1 \approx 1$, and vanishes in the parameter regimes $J_0 \gtrsim 2$ and $J_1 \gtrsim 4$.
It is interesting to note that for some values in the region $J_{0}<1$ and $J_{1}>1$, the asymptotic discord has finite values whereas entanglement vanishes \cite{Sadiek_2010} for all values in this region.
Fig.~\ref{[J(),h()]-T0-gamma1}(b) is the discord contour of $h_{0}$ and $h_{1}$ with $J_{0}=J_{1}=1$. The discord starts with zero at $h_1=h_0=0$ and reaches the maximum at $h_{0} \approx h_{1} \approx 1$.
We then plot the contour $D(i,i+1)$ as a function of $J_{0}$ and $h_{0}$ in Fig.~\ref{[J(),h()]-T0-gamma1}(c).
The discord has its maximum when $J_{0}\approx h_{0}$ and disappears when $J_{0}$ deviates largely from $h_{0}$. We also show the asymptotic behavior of $D(i,i+1)$ as a function of $J_{1}$ and $h_{1}$
 as shown in Fig.~\ref{[J(),h()]-T0-gamma1}(d). The largest discord is reached at $J_{1}=h_{1}$.
The profile of the discord contour versus $J_0$ and $h_0$ is smooth, whereas the profile versus $J_{1}$ and $h_{1}$ is not. The reason is that $J_{0}$ and $h_{0}$ only affect the system at the initial time, whereas $J_{1}$ and $h_{1}$ affect the system all the time. A little change in the values of $J_{1}$ and $h_{1}$ has a great impact on the asymptotic behaviors of quantum discord. Fig.~\ref{[J(),h()]-T0-gamma0.5} is the same as Fig.~\ref{[J(),h()]-T0-gamma1} except $\gamma=0.5$.
The asymptotic behavior of discord is similar to that of the Ising case as shown in Fig.~\ref{[J(),h()]-T0-gamma0.5}.
The differences are that the quantum discord does not vanish in the range of $J_{0}>1$ and $J_{1}>1$, and the maxima of quantum discord in the partially anisotropic $XY$ model are smaller than those in the Ising model. There is no peak in the vicinity of $J_{1}=h_{1}$.
Fig.~\ref{[J(),h()]-T0-gamma0} is for the isotropic $XY$ model, where $D(i,i+1)$ depends only on the $J_{0}$ and $h_0$ but not on $J_{1}$ and $h_1$.

The temperature effect on quantum discord is shown in Fig.~\ref{(kT,lambda())-gamma1}. In Fig.~\ref{(kT,lambda())-gamma1}(a), the quantum discord reaches the maximum at $\lambda=1$.
As the temperature increases or $\lambda$ diverges from the critical value, the quantum discord decays more slowly than entanglement \cite{Sadiek_2010}. The asymptotic behaviors of $D(i,i+1)$ as a function of $\lambda_{1}$, $\lambda_{0}$ and $kT$, are depicted in Fig.~\ref{(kT,lambda())-gamma1}(b) and \ref{(kT,lambda())-gamma1}(c). These figures demonstrate that the quantum discord is more robust than entanglement against the effect of temperature.


\begin{figure*}[htbp]
\centering
   \subfigure{\label{(kT,lambda())-gamma1}\includegraphics[width=15cm]{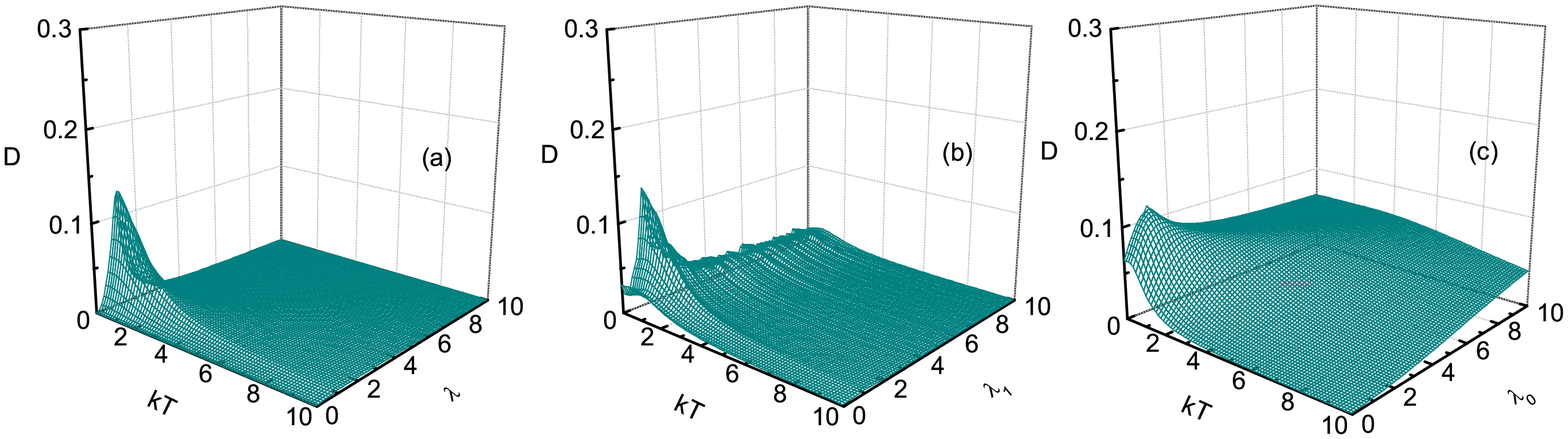}}
 \caption{(Color online)The asymptotic behavior of $D(i,i+1)$ as a function of $kT$ and (a)$\lambda$  with $J_{0}=J_{1}$, (b) $\lambda_{1}$ with $J_{0}=1$, (c) $\lambda_{0}$, and $J_{1}=1$. $h_{0}=h_{1}=1$. $\gamma=1$.}
 \label{(kT,lambda())-gamma1}
\end{figure*}


Finally, we explore the asymptotic behaviors of quantum discord in the $\lambda_{1}$-$\gamma$ phase space for the different values of $J_0$. In Figs.~\ref{(gamma,lambda())-T0}(a), \ref{(gamma,lambda())-T0}(b) and \ref{(gamma,lambda())-T0}(c),  we plot $D(i,i+1)$ as a contour function of $\lambda_1$ and $\gamma$ for given magnetic fields $h_{0}=h_{1}=1$ at $kT=0$.
When $J_0 <1$, there is no bigger enhancement in $D(i,i+1)$.
When $J_0$ is in the vicinity of $1$, there are two peaks. For the larger $J_0$, two peaks merge.
When $J_0 >1$ and $\lambda_1<1$, $D(i,i+1)$ is enhanced significantly in the $\gamma=0$ region, whereas for $\lambda_1>1$, $D(i,i+1)$ is not enhanced.


\begin{figure*}[htbp]
\centering
   \subfigure{\label{(gamma,lambda())-T0}\includegraphics[width=15cm]{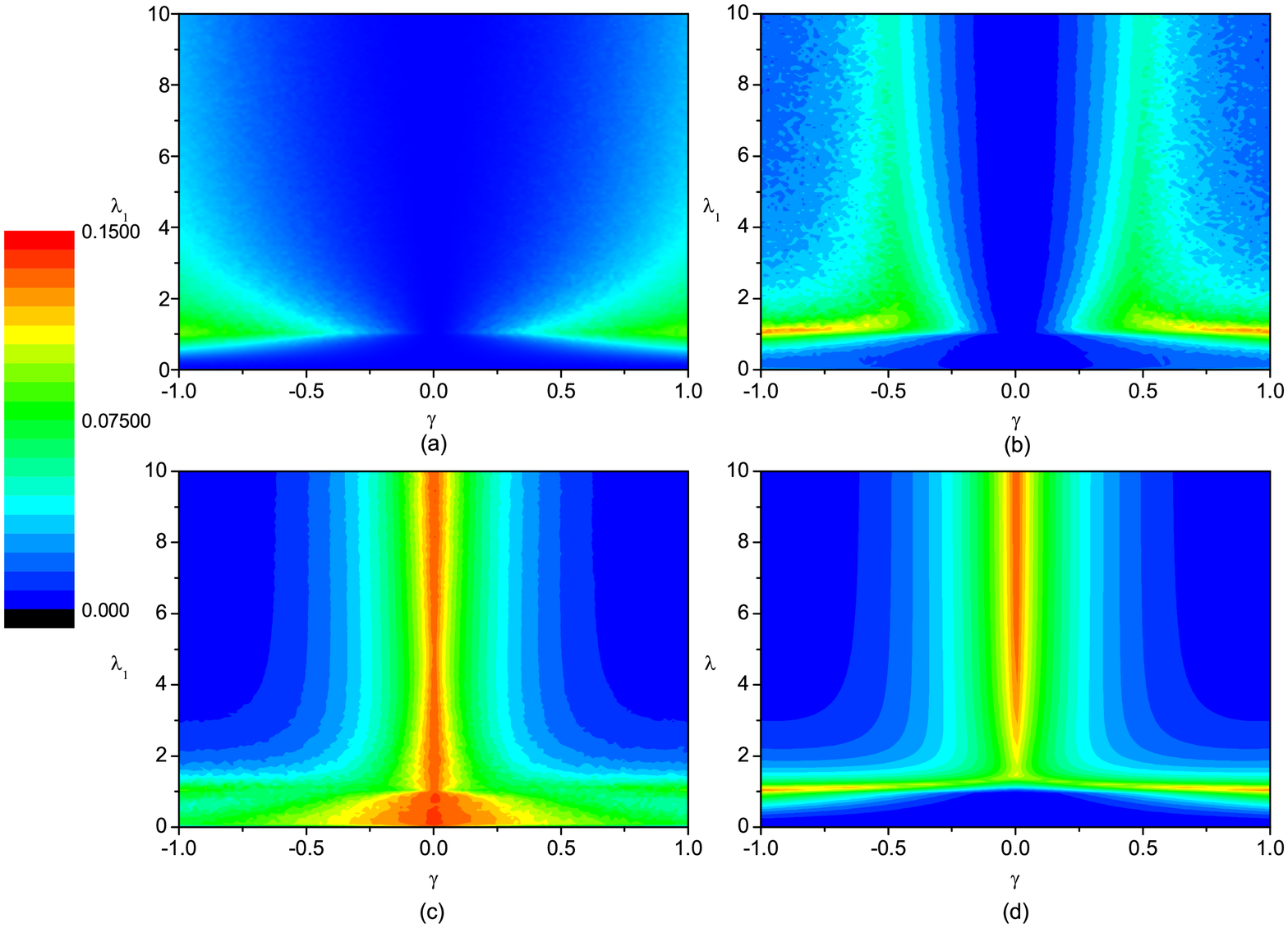}}
   \caption{(Color online) The asymptotic behavior of $D(i,i+1)$ as a function of (a)$\lambda_{1}$ and $\gamma$ with  $J_{0}=0.5$, (b)$\lambda_{1}$ and $\gamma$ with $J_{0}= 1$, (c)$\lambda_{1}$ and $\gamma$ with $J_{0}=5$, and (d)$\lambda$ and $\gamma$ at $kT=0$. $h_{0}=h_{1}=1$.}
 \label{(gamma,lambda())-T0}
\end{figure*}


\section{Conclusions}
We have investigated the dynamics and the asymptotic behaviors of quantum discord in one dimensional $XY$ model coupled through a time-dependent nearest-neighbor coupling and in the presence of a time-dependent magnetic field.  The system shows non-ergodic and critical behaviors. The zero-temperature asymptotic behaviors of the system at the long time limit depends only on the ratio between the coupling and the magnetic field but not their individual values. On the contrary, these behaviors do rely on individual values of the coupling and the magnetic field at the finite temperature.
The asymptotic behaviors are sensitive not only to the initial values of the coupling and the magnetic field, but also to the final values of the coupling and the magnetic field. We have demonstrated that quantum discord is more robust than entanglement against the effect of temperature. This robustness will be useful in design of fault-tolerant quantum algorithms and in other quantum information processing.

\section*{Acknowledgment}
We acknowledge the financial support by the National Natural Science Foundation of China under Grant No. 11075013 and No. 10974016.
L.-A. Wu has been supported by the Ikerbasque Foundation Startup, the Basque Government (Grant IT472-10), and the Spanish MEC(Project No. FIS2009-12773-C02-02).


\end{CJK*}
\end{document}